\documentclass[pre,twocolumn,superscriptaddress,aps,nofootinbib]{revtex4-1}
\usepackage{stix}
\usepackage{empheq}
\usepackage{amsmath}
\usepackage{tabularx}
\usepackage{dsfont}
\usepackage{graphicx}
\usepackage{hyperref}
\usepackage{stackrel}
\usepackage{amssymb,latexsym,mathrsfs}
\usepackage{amsfonts}
\usepackage{srcltx}
\usepackage{color}
\usepackage{cancel}
\usepackage{xr}

\def\d{{\rm d}}
\def\0{\emptyset}

\begin{document}
	
	\title{Diffusion and first-passage characteristics on a dynamically evolving support}
	\author{Manuel Schrauth}
	\email{manuel.schrauth@uni-wuerzburg.de}
	\affiliation{Institute of Theoretical Physics and Astrophysics,	University of W\"urzburg, 97074 W\"urzburg, Germany}

	\author{Maximilian Schneider}
	\affiliation{Institute of Theoretical Physics and Astrophysics,	University of W\"urzburg, 97074 W\"urzburg, Germany}
	\affiliation{Department of Informatics, Technical University of Munich, 85748 Garching, Germany}

\begin{abstract}
	We propose a generalized diffusion equation for a flat Euclidean space subjected to a continuous infinitesimal scale transform. For the special cases of an algebraic or exponential expansion/contraction, governed by time-dependent scale factors $ a(t)\sim t^\lambda $ and $ a(t)\sim\exp(\mu t) $, the partial differential equation is solved analytically and the asymptotic scaling behavior, as well as the dynamical exponents, are derived. Whereas in the algebraic case the two processes (diffusion and expansion) compete and a crossover is observed, we find that for exponential dynamics the expansion dominates on all time scales. For the case of contracting spaces, an algebraic evolution slows down the overall dynamics, reflected in terms of a new effective diffusion constant, whereas an exponential contraction neutralizes the diffusive behavior entirely and leads to a stationary state. Furthermore, we derive various first-passage properties and describe four qualitatively different regimes of (strong) recurrent/transient behavior depending on the scale factor exponent.
\end{abstract}	
	
\maketitle

\section{Introduction}
\label{sec:Introduction}

	A cornerstone of our understanding of critical phenomena is the concept of self-similarity and scale invariance~\cite{fisher1974,wilson1975,kadanoff-cardy}. These terms refer to a situation in which the physical state of a system is in some sense invariant under a change of scale of the supporting geometry. Scale-invariant properties are often found to be universal, i.e.~they depend only on the symmetries of the system but not on the specific microscopic realization.
	
	In most studies, scale invariance is used as a mathematical tool for the analysis of critical phenomena. The aim of the present study is to investigate scale transformations from a different perspective, namely, as part of the physical process itself. More specifically, we consider time-dependent processes in which infinitesimal scale transformations are continuously carried out as part of the dynamics. As an example, this would correspond to a self-inflating (or self-deflating) supporting geometry on which the process takes place. If the process itself is scale-invariant, it is interesting to study how it responds to the continual change of scale of the underlying support. As a possible motivation, such self-inflating scale-free processes may be regarded as toy models for dynamical phenomena taking place in an expanding universe and in particular for the early phase of cosmological inflation~\cite{liddle2009}. Indeed, very recently a couple of studies investigated a massless gas in a homogeneously and isotropically expanding space, having cosmological expansion in view \cite{bazow2016a,bazow2016b}. Furthermore, \cite{camilo2016} studies a strongly coupled conformal field theory plasma also subject to an expanding Friedmann-Lema\^{i}tre-Robertson-Walker metric. Apart from that, our considerations may serve as a basis to better understand self-similar processes taking place on (biological) growing substrates or surfaces \cite{barabasi1995}. 
	
	\begin{figure}
		\centering\includegraphics[width=\linewidth]{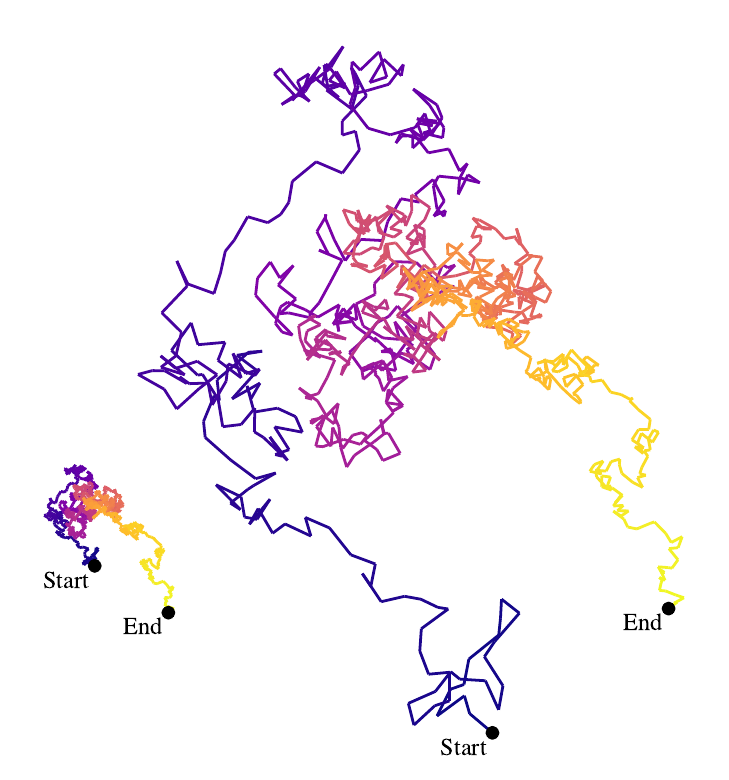}
		\caption {(Color online) Left: Ordinary random walk on a static plane. The changing color scale indicates the advance of time. Right: The corresponding random walk with the same sequence of microscopic displacements taking place on a self-inflating plane. Shown is the situation at the end of the simulation. As can be seen, earlier steps have undergone a stronger expansion compared to later ones.}
		\label{fig:rw}
	\end{figure}	
	
	In order to gain an appreciation of how background expansion affects the dynamics of diffusing particles, let us consider an ordinary two-dimensional random walk of static step length, as shown on the left hand side of Fig.~\ref{fig:rw}. If the same random walk were placed on a homogeneously expanding background, the resulting path would be different. For example, in the case of an exponential expansion, this could be realized by stretching all segments of the path by a factor $ s $ after each move. The resulting snapshot of the path is shown on the right hand side of Fig.~\ref{fig:rw}. Illustratively stated, this could be the path of an ant on an inflating balloon. Obviously, early moves have been stretched over time while recent moves still have approximately the original size. Note that the inflation leads to an entirely different path although the sequence of steps is the same in both cases.
	
	The figure also suggests that the same path could have been generated by a random walk on a \textit{static} background but with a dynamically shrinking step size. For the particular case of an exponential expansion these systems are called Bernoulli convolution in the mathematical literature and have already been studied since the 1930s~\cite{kershner1935,erdoes1939}. For the case of exponentially shrinking (growing) steps mentioned before, the random walk has also been investigated from a more physical perspective in a number of studies \cite{delatorre2000,krapivsky2003,subramanian2015}, although the correspondence to an exponentially growing (shrinking) support is not addressed there. Instead, the support is treated as static and the shrinking step size is supposed to be caused by some physical mechanism.  In particular, starting with a step size of $s^0 = 1$, the length of the $n$-th step is given by $s^{n-1}$. It is immediately obvious that for $s=1$ the ordinary random walk is recovered, whereas the steps are shrinking (growing) for smaller (larger) values. 
	
	It turns out that for this exponential setting the discrete random walk exhibits a number of appealing new features. Most intriguing is the spatial probability density $ p_s(x) $, which shows a fractal structure for $ s<1/2 $, a uniform probability density for $ s = 1/2 $, and is again fractal if $ s > 1/2 $. In particular, for $ s = (\sqrt{5}-1)/2 $ (the inverse of the golden ratio) the distribution becomes strikingly self-similar \cite{krapivsky2003} (see e.g.~Fig.~5 in Ref.~\cite{krapivsky2003}). Finally, when $ s \rightarrow 1$ the probability density becomes smoother and ultimately approaches a Gaussian, as expected in the limit of non-shrinking steps or a non-expanding background, respectively. Besides these interesting results, also first-passage properties \cite{rador2006a} and multidimensional settings \cite{rador2006b,serino2010} have been considered for discrete random walks with exponentially varying step sizes. 
	
	In contrast to the aforementioned results, which were mostly based on numerical simulations of a discrete random walker model, we focus in the following on the continuous formulation of the diffusion problem. Therefore, we derive an analytical diffusion equation on a dynamical background and address the two particular cases of an exponential and algebraic inflation/deflation explicitly. Furthermore, the continuous formulation allows us to derive the first-passage behavior of our setting to some extend analytically.
	
	The outline of the paper is as follows. In Section II we introduce our basic notation. Section III presents our main equation, combining inflation/deflation and diffusion, as well as its solution for algebraic and exponential dynamics. Also the generalization to higher dimensions is mentioned briefly. In Section IV, corresponding first-passage characteristics are discussed. A summary of the results is given in Section V.

\section{Basic notation}
	\label{sec:BasicNotation}
	
	We consider a Euclidean space which is expanding homogeneously and isotropically by itself. Choosing a given point as the origin, any other point, with position $x(t)\in\mathbb{R}^d$ will move away from the origin. More specifically, the corresponding position vector $ x(t) $ changes according to the differential equation
	\begin{align}
		\label{eq:definition_evolving_space}
		\d x(t)=H(t)\,x(t)\, \d t,
	\end{align}
	where the scalar-valued quantity $H(t)$ is the time-dependent expansion rate. Alluding to cosmology, we will refer to it as the \emph{Hubble parameter}. For a given Hubble parameter, we can define the dimensionless scale factor $a(t)$. The scale factor itself evolves by the differential equation
	\begin{align}
		H(t)=\frac{\dot a(t)}{a(t)}
	\end{align}
	with formal solution
	\begin{align}
		a(t) \;=\; a(t_0) \,\exp \left[ \int_{t_0}^t \d t'\, H(t') \right].
		\label{eq:scalefactor}
	\end{align}
	It tells us by which factor the length scales have been inflated\footnote{Unless otherwise stated, we use the terms \textit{inflation}, \textit{expansion} etc.~as a shorthand expression for \textit{inflation/deflation}, \textit{expansion/contraction}, etc. throughout this article.} with respect to a certain reference time $t_0$. Since it is convenient to start from a non-stretched system, we set $ a(t_0)=1 $.

\section{Diffusion}
\label{sec:Diffusion}
	
	In order to derive a differential equation for diffusion on a background that is evolving in time, we start from a microscopic description, namely from one-dimensional Brownian motion \cite{lemons2002} for a single particle and add the expansion term from Eq.~\eqref{eq:definition_evolving_space}, i.e.,
	\begin{equation}
		\d x(t)=H(t)\,x(t)\, \d t+\zeta(t)\d t
		\label{eq:langevin1D}
	\end{equation}
	where $\zeta(t)$ denotes Gaussian white noise with zero mean $\left< \zeta(t) \right>=0$ and correlations $\left< \zeta(t) \zeta(t')\right>=2D\, \delta(t-t')$. $D$~is the diffusion constant. It is now straightforward to derive the corresponding Fokker-Planck equation for the probability density $p(x,t)$ of many non-interacting diffusing particles, e.g.~by It\^o calculus or Kramers-Moyal expansion~\cite{gardiner1985}, giving the partial differential equation
	\begin{equation}
		\partial_t\, p(x,t)=-H(t)\, \partial_x \Big(x\ p(x,t) \Big) +D\, \partial_x^2\, p(x,t),
		\label{eq:inflation_diffusion_equation}
	\end{equation}
	where the first term on the right hand side accounts for the homogeneous and isotropic expansion of the underlying space and the second part describes ordinary diffusion on top of it. As can be easily verified, the normalization of the probability density $p$ is conserved in time, so the additional term accounting for the background dynamics is in this respect well behaved and introduces no gain or loss of particle density to the system. We refer to Eq.~\eqref{eq:inflation_diffusion_equation} by the name \emph{inflation-diffusion equation}, because in the remaining part of this paper we only deal with the special case of an inflating (or deflating) support. % However, it should be remarked here that Eq.~\ref{eq:inflation_diffusion_equation} is fairly general and \red{holds for \textit{any} time-dependent background dynamics}.
	
	In particular we consider two different choices for the Hubble parameter $H(t)$. First, if $H(t)=\mu$ is a constant, the scale factor reads 
	\begin{equation}
		a(t)=\exp\left(\mu(t-t_0)\right),\qquad\qquad t\geq t_0
		\label{eq:exponentialgrowth}
	\end{equation}
	which describes an exponentially driven expansion ($\mu\!>\!0$) or contraction ($\mu\!<\!0$) of space. The absolute value of $\mu$ is therefore a measure of the strength of the exponential dynamics, or, stated differently, denotes the inverse of the expansion/contraction time scale $\tau\sim\mu^{-1}$. We set the initial time, $t_0$, equal to zero for convenience. 
	
	The second choice of $H(t)$ we consider in this article corresponds to a background with algebraic dynamics, i.e. the evolving scale factor is characterized by a power-law behavior 
	\begin{equation}
		a(t)=\left(\frac{t}{t_0}\right)^\lambda,\qquad\qquad t\geq t_0
		\label{eq:algebraicgrowth}
	\end{equation}
	where the dimensionless parameter $\lambda$ again accounts for the strength of the inflation/deflation. Once more, $t_0$ denotes the initial time, which we set to $t_0=1$ in this setting. The corresponding Hubble parameter is given by $H(t) = \lambda/t$ in this case. 
	
	As already mentioned in Sec.~\ref{sec:Introduction}, the random walk on an expanding (contracting) space is equivalent to the same walk on a static background but with shrinking (growing) step size. For the special case of exponential inflation constant $H>0$, the inflation-diffusion equation~\eqref{eq:inflation_diffusion_equation} is therefore equivalent to the continuum equation for a random walker with exponentially shrinking step size, as proposed by Rador and Taneri~\cite{rador2006a} where the diffusion constant was made time dependent, $D(t) = D_0 \exp(-t/\tau)$, with some time scale $\tau$. This equivalence can be shown by transforming Eq.~\eqref{eq:inflation_diffusion_equation} back into non-expanding coordinates 
	\begin{equation}
		(x,t)\rightarrow(s,t') \equiv \Big(a^{-1}(t)\,x,\,t\Big),
	\end{equation}
	taking proper care of derivatives, which in this case transform as
	\begin{eqnarray}
		\partial_x = \frac{\partial t'}{\partial x}\partial_{t'} + \frac{\partial s}{\partial x}\partial_s = a^{-1}(t')\,\partial_s\\
		\partial_t = \frac{\partial t'}{\partial t}\partial_{t'} + \frac{\partial s}{\partial t}\partial_s = \partial_{t'} - H(t')s\,\partial_s,
	\end{eqnarray}
	then inserting the scale factor and Hubble parameter corresponding to exponential inflation, Eq.~\eqref{eq:exponentialgrowth}, and, finally, identifying $\mu \rightarrow 1/2\tau$.
	
	\subsection{Characteristics for algebraic dynamics}
	\label{sec:CharacteristicsAlgebraical}
	
	In order to study the qualitative behavior of the inflation-diffusion equation for the case of the algebraic dynamics as described by Eq.~\eqref{eq:algebraicgrowth} we solve it on an infinite line with initial condition $p(x,t=1)=\delta(x-x_0)$. This can be achieved by transforming the spatial coordinate into Fourier space,
	\begin{align}
		p(x,t)\rightarrow \hat{p}(k,t)=\frac{1}{\sqrt{2\pi}}\int\limits_{-\infty}^{\infty}\d x\,p(x,t)\,\mathrm{e}^{ikx},
	\end{align}
	such that Eq.~\eqref{eq:inflation_diffusion_equation} becomes
	\begin{equation}
		\partial_t\,\hat{p}(k,t)= H(t)\,k\, \partial_k\,\hat{p}(k,t)-D\, k^2\, \hat{p}(k,t).
		\label{eq:fokker1Dfourier}
	\end{equation}
	Inserting $H(t)=\lambda/t$, the solution is readily given by
	\begin{equation}
		\hat{p}(k,t)=e^{-\frac{D\, k^2\,t}{1-2\lambda}} f(k\,t^\lambda)
	\end{equation}
	where $f$ represents some scaling function which is determined by the initial condition. Using an initial delta peak at $x=x_0$ as stated above we arrive at 
	\begin{equation}
		\hat{p}(k,t)=\frac{1}{\sqrt{2\pi}} \exp\left(-\frac{1}{2}k^2\xi_\lambda^2 + ik\,\langle x\rangle_\lambda\right)
		\label{eq:fokker1DpowFT}
	\end{equation}
	where we introduced
	\begin{equation}
		\xi_\lambda \equiv \sqrt{\big\langle(x-\left<x\right>_\lambda)^2\big\rangle} = \sqrt{\frac{2D(t-t^{2 \lambda })}{1-2 \lambda}}
	\end{equation}
	and
	\begin{equation}
		\langle x\rangle_\lambda \equiv x_0t^\lambda
	\end{equation}
	as shorthand notations for the standard deviation of the spatial displacement and its mean value, respectively, where the time dependence is left implicit. The Fourier space solution can be transformed back analytically to yield
	\begin{equation}
		p(x,t)=\frac{1}{\sqrt{2 \pi}\xi_\lambda} \exp \left[-\frac{\left(x-\langle x\rangle_\lambda\right)^2}{2\,\xi_\lambda^2}\right],
		\label{eq:pdf1Dpow}
	\end{equation}
	Clearly, this expression arises from the fundamental solution of the ordinary diffusion equation in static space by the following substitutions 
	\begin{equation}
		t\rightarrow \frac{t-t^{2\lambda}}{1-2\lambda},\qquad
		\langle x \rangle =x_0\rightarrow \langle x\rangle_\lambda=x_0\,t^\lambda.
		\label{eq:SubstitutionsAlgebraical}
	\end{equation}
	Note that for $ \lambda<0 $, i.e.~a contraction of space, the diffusion is not stopped but only slowed down by a factor of $ 1-2\lambda $, which can be seen from \eqref{eq:SubstitutionsAlgebraical} for $ t\rightarrow\infty $. This can also be expressed through an effective diffusion constant
	\begin{align}
		D_\text{eff}(\lambda) = \dfrac{D}{1-2\lambda}.
	\end{align}
	The preceding calculations enable us to determine the dynamical exponent $z$, given by the asymptotic relation $\xi\sim t^{1/z}$. In particular, we need to distinguish three different cases in the long-time limit $t\rightarrow\infty$
	\begin{equation*}
		\lim_{t\rightarrow \infty} \xi_\lambda \sim \left\{\begin{array}{cl}
			\sqrt{t} \quad      &\mbox{for}\quad \lambda<1/2,\\
			\sqrt{t\log t}\quad &\mbox{for}\quad \lambda=1/2,\\
			t^\lambda\quad      &\mbox{for}\quad \lambda>1/2.
		\end{array}\right.\,
	\end{equation*}
	Hence, there exists a critical value of $\lambda_c = 1/2$ that subdivides the dynamics of the system into two different regimes corresponding to the two competing terms in the inflation-diffusion equation. For $\lambda > 1/2$ inflation overwhelms the ordinary diffusive behavior, whereas for $\lambda < 1/2$ the diffusion term dominates the evolution of the system. Precisely at the critical $\lambda_c$, the system shows ordinary diffusive behavior but is subject to logarithmic corrections due to the expansion term.
	
	The dynamical exponent is therefore given by
	\begin{equation*}
		z= \left\{\begin{array}{cl}
			2\quad &\mbox{for}\quad \lambda\leq 1/2, \\
			\lambda^{-1}\quad &\mbox{for} \quad\lambda> 1/2.
		\end{array}\right.\,
	\end{equation*} 
	
	\subsection{Characteristics for exponential dynamics}
	
	We now turn to the case of exponential dynamics where the Hubble parameter is given by $H(t)=\mu$. Using again the transformed Eq.~\eqref{eq:fokker1Dfourier}, the solution for initial conditions $p(x,t=0)=\delta(x-x_0)$ can be obtained in the same manner as for the algebraic dynamics and reads
	\begin{equation}
		p(x,t)=\frac{1}{\sqrt{2 \pi}\xi_\mu} \exp \left(-\frac{\left(x-\langle x\rangle_\mu\right)^2}{2\,\xi_\mu^2}\right),
		\label{eq:pdf1Dexp}
	\end{equation}
	where the width and mean of the distribution are now given by 
	\begin{equation}
		\xi_\mu \equiv \sqrt{\left<(x-\left<x\right>_\mu)^2\right>} = \sqrt{\frac{D}{\mu}\left(\mathrm{e}^{2\mu t}-1\right)} 
		\label{eq:StD-exponential}
	\end{equation}
	and
	\begin{equation}
		\langle x \rangle_\mu\equiv x_0\,e^{\mu t},
	\end{equation}
	respectively. Again the solution can be obtained from the ordinary heat kernel by the substitutions
	\begin{equation}
		t\rightarrow \frac{e^{2\mu t}-1}{2\mu}, \qquad
		\langle x \rangle \rightarrow \langle x \rangle_\mu.
	\end{equation}
	Regarding Eq.~\eqref{eq:StD-exponential} it is obvious that for an \textit{expansion} of the system, i.e.~$\mu>0$, the width of the distribution scales exponentially for large times $\xi_\mu\sim\mbox{e}^{\,\mu t}$ and therefore a dynamical exponent can not be defined here. Roughly speaking, the exponentially driven inflation is so strong that the diffusive behavior is completely overwhelmed. 
	
	However, the picture is different when $\mu<0$, corresponding to an exponentially \textit{shrinking} support. In this case, Eq.~\eqref{eq:pdf1Dexp} eventually approaches a stationary state, given by a Gaussian distribution with fixed width 
	\begin{equation}
		\lim_{t\rightarrow\infty}\xi_{\mu<0}=\sqrt{\frac{D}{|\mu|}}
		\label{eq:fixed_width}
	\end{equation}
	This means that exponential deflation of the underlying space in some sense neutralizes the diffusive behavior, as the width of the stationary state is simply given by the ratio of the two constants that measure the strength (or time scale) of each process.

	\subsection{Generalization to higher dimensions}
	
	For the sake of completeness, we also present the multi-dimensional form of the inflation-diffusion equation
	\begin{align}
		\frac{\partial}{\partial t} p=-\mathrm{div} \Big(H\,x\,p\Big) + D\,\mathrm{div}\Big(\mathrm{grad}\,p\Big),
		\label{eq:fokkerHDtensor}
	\end{align}
	where $ x $ now denotes a $ d $-dimensional vector, $ p\equiv p(x,t) $
	and $ H $ is in the most general case, i.e.~for an arbitrary dynamics of the underlying space, defined as a $d\!\times\!d$-tensor with components depending on both $ x $ and $ t $. For homogeneous and isotropic expansion or contraction, $H$ reduces to a time-dependent scalar and Eq.~\eqref{eq:fokkerHDtensor} reads
	\begin{multline}
		\frac{\partial}{\partial t} p(x,t)= -H(t)\left(d+\sum_{i=1}^{d}x_i\frac{\partial}{\partial x_i}\right) p(x,t)\\
		+D\sum_{i=1}^{d}\frac{\partial^2}{\partial x_i^2} \, p(x,t)
		\label{eq:fokkerHDcart2}
	\end{multline}
	in $d$-dimensional Cartesian coordinates and 
	\begin{multline}
		\frac{\partial}{\partial t} p(r,t)=-H(t)\frac{1}{r^{d-1}}\frac{\partial}{\partial r}\left(r^d p(r,t)\right) +\\
		D\,\frac{1}{r^{d-1}}\frac{\partial}{\partial r}\left(r^{d-1}\frac{\partial}{\partial r}p(r,t)\right),
		\label{eq:sphericalFokker}
	\end{multline}
	for the special case of spherical coordinates, with $r=\sqrt{x_1^2+x_2^2+\dots+x_d^2}$. It turns out that the isotropic radial equation \eqref{eq:sphericalFokker} can be solved analytically for both choices of the Hubble parameter we are considering in this paper. The procedure is similar to the one-dimensional case, with the difference being that now a Hankel transform needs to be employed instead of the ordinary Fourier transform. We refer the reader to \cite{duffy} where the procedure is explained in some detail. The initial condition is now given by a $d$-dimensional spherical shell at position $r=r_0$
	\begin{equation}
		p (r, t = 1) = \frac {1} {r_ 0^{d - 1}\Omega_d}\delta (r - r_ 0)
		\label{eq:RadialInitialCondition}
	\end{equation}
	where $\Omega_d=2\pi^{\frac{d}{2}}/\Gamma(\frac{d}{2})$ denotes the surface area of the $d$-dimensional unit sphere. Eventually, we arrive at
	\begin{multline}
		p(r,t)=\frac{1}{\Omega_d}\frac{\left(r\, \langle r\rangle_\alpha\right)^{1-d/2}}{\xi_\alpha^2}\times \\
		\exp \left(-\frac{r^2+\langle r\rangle_\alpha^2}{2\,\xi_\alpha^2} \right)
		I_{d/2-1}\left(\frac{r\, \langle r\rangle_\alpha}{\xi_\alpha^{2}}\right)
		\label{eq:pdf_HD}
	\end{multline}
	where $I_\nu$ represents the \textit{modified Bessel function of the first kind} and of order $\nu$ \cite{abramowitz1972} and the index $\alpha\in\{\lambda,\mu\}$. For $r_0=0$, i.e.~for diffusing particles that start at the origin, the Bessel function vanishes and we recover a standard Gaussian form.

\section{First-passage properties}
\label{sec:FirstPassageProperties}
	
	After having investigated the behavior of diffusion on a dynamical background, we are now going to discuss the associated first-passage properties. First-passage processes appear in the context of various phenomena in nature, such as the firing of neurons \cite{gerstein,fienberg,tuckwell} or the initiation of chemical reactions \cite{rice}. They are also interesting on their own, as the scaling behavior is known to show a non-trivial dependence on the dimensionality. 
	
	A diffusing particle in one-dimensional static space is \emph{recurrent}, which means that it is certain to eventually passage any site. However, on average, it takes an infinitely long time to arrive. Whereas for two dimensions recurrence is still valid, the situation changes for higher dimensions, as some sites are never being visited by the particle on its random walk. Consequently, only a fraction of particles passages the origin. This is called \emph{transience}. In the following we are particularly interested in how the main first-passage characteristics, given by the survival probability and the mean first-passage time, change if the background is evolving, as defined in Sec.~\ref{sec:Diffusion}. Moreover, we investigate how the first-passage rate, which in the one-dimensional static space declines according to the power law $t^{-3/2}$, is affected by the dynamical underlying space.
	
	\subsection{Image method}
	
	The first-passage problem for a diffusing particle concentration $c(x,t)$ is directed by the inflation-diffusion equation~\eqref{eq:inflation_diffusion_equation}, with the additional boundary condition 
	\begin{align}
		c(x=0,t)=0.
		\label{eq:AbsorbingBoundaryConditions}
	\end{align}
	This accounts for the fact that each particle leaves the system as soon as it reaches the origin. Due to this absorbing sink, the overall particle concentration is not constant but decreases with time. Hence, it has to be remarked that $c(x,t)$ is not a probability density in the strict sense, as its norm is always smaller than one for $t>t_0$.
	
	A common method to solve such a Dirichlet boundary problem involves applying a Fourier sine transform defined by ${\hat{c}(k,t)=\int_0^\infty \d x\, c(x,t)\, \sin(k\,x)}$ to the governing differential equation. However, since we have already calculated the general solution on the open domain, we simply use an image method. More specifically, we obtain the solution of the first-passage problem for a particle starting somewhere in the positive half-space  ${x=x_0>0}$ by superposing the probability density $p_+(x,t)$ with the negative density $p_-(x,t)$ of a hypothetical image particle with initial position $(-x_0)$. The solution $c(x,t)=p_+(x,t)+p_-(x,t)$ fulfills both the inflation-diffusion equation~\eqref{eq:inflation_diffusion_equation} and the boundary condition~\eqref{eq:AbsorbingBoundaryConditions}.
	
	The probability density of the particle and its image particle in algebraically or exponentially evolving spaces are given by
	\begin{equation}
		p_{\pm}(x,t)=\pm \frac{1}{\sqrt{2 \pi} \xi_{\alpha}} \exp \left[-\frac{1}{2}  \frac{\left(x\mp \langle x\rangle_{\alpha}\right)^2}{\xi_{\alpha}^2}\right],
	\end{equation}
	respectively, where again $\alpha\in\{\lambda,\mu\}$.  The full solution therefore reads
	\begin{equation}
		c(x,t)=\frac{1}{\xi_{\alpha}}\sqrt{\frac{2}{\pi}} \exp \left(-\frac{1}{2}\frac{x^2+\langle x \rangle_{\alpha}^2 }{\xi_{\alpha}^2}\right) \sinh \left(\frac{x\, \langle x \rangle_{\alpha} }{\xi_{\alpha}^2} \right).
		\label{eq:FullSolutionImageMethod}
	\end{equation}
	This concentration of particles that have not yet passaged the origin contains all the information required to derive the first-passage properties. The first-passage rate $F(t)$, for instance, is nothing else than the flux of particles to the origin,
	\begin{align}
		F(t)\equiv D\frac{\partial}{\partial x}c(x,t)\big|_{x=0}.
		\label{eq:FirstPassageRate}
	\end{align}
	
	\subsection{Characteristics for algebraic dynamics}
	
	Let us again turn specifically to an algebraically inflating or deflating space. In this case the first-passage rate is found to scale as
	\begin{equation}
		F(t)\sim \frac{t^\lambda}{\xi_\lambda^3}\sim \left(\frac{1-2\lambda}{t^{1-\frac{2}{3} \lambda}-t^{\frac{4}{3}\lambda}}\right)^{3/2},
		\label{eq:FPR_Scaling_Algebraical}
	\end{equation}
	according to Eqs.~\eqref{eq:FullSolutionImageMethod} and \eqref{eq:FirstPassageRate}. In the asymptotic limit of large times we therefore get a power law behavior $F(t)\sim t^{\kappa}$, with the so-called first-passage exponent $\kappa$. Following Eq.~\eqref{eq:FPR_Scaling_Algebraical}, it is given by
	\begin{equation}
		\kappa=\left\{\begin{array}{cl}
			-\frac{3}{2}+\lambda & \quad\mbox{ if}\quad \lambda\leq1/2, \\
			-2\lambda & \quad\mbox{ if}\quad \lambda>1/2
		\end{array}\right. \quad \mathrm{for}\quad t \rightarrow \infty.
		\label{eq:kappa}
	\end{equation}
	\begin{figure}
		\centering\includegraphics[width=\linewidth]{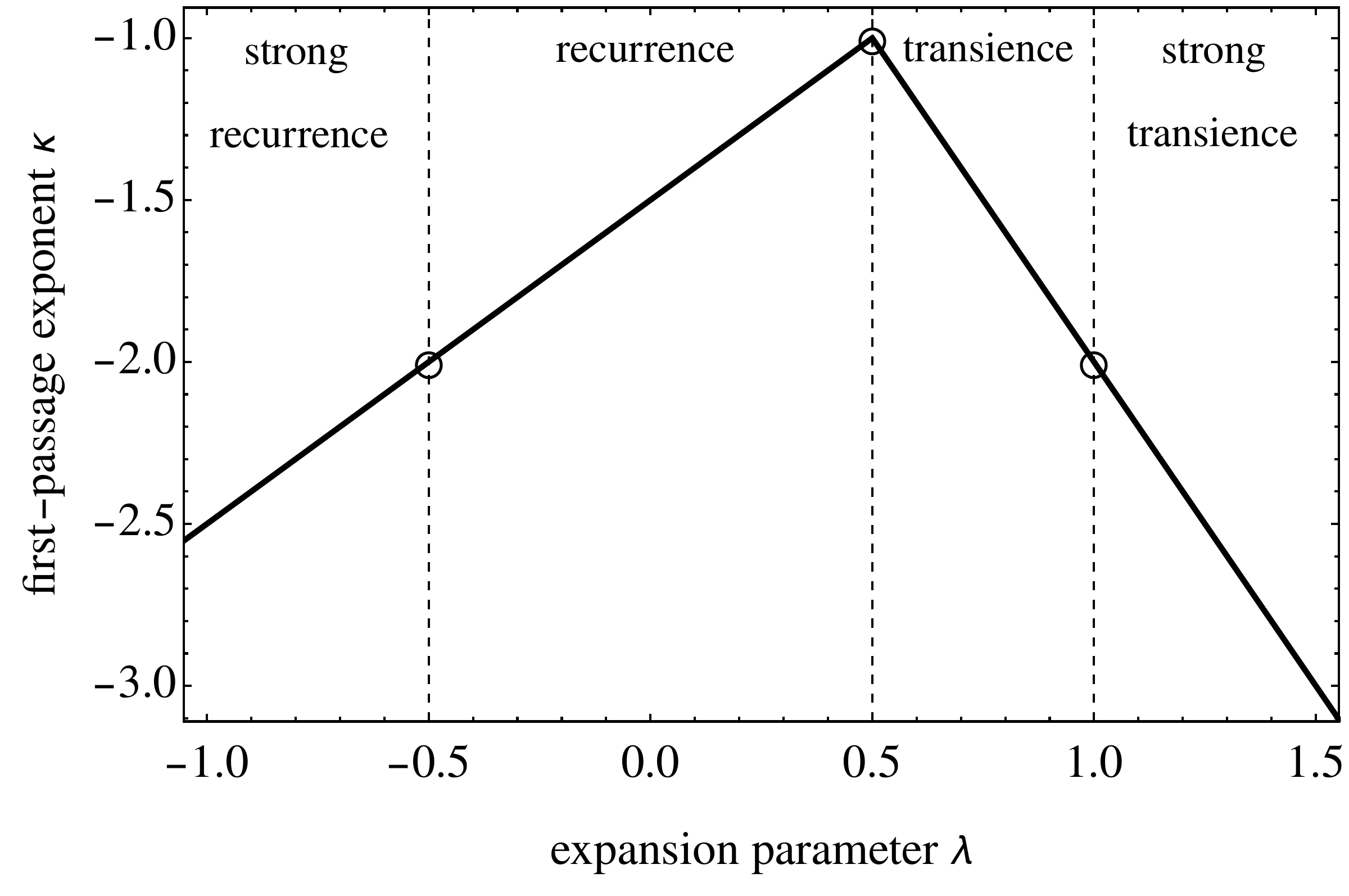}
		\caption{The first-passage exponent $\kappa$ as a function of the algebraic expansion parameter $\lambda$. The dashed lines separate four regions of qualitatively different behavior which are explained in the text.}
		\label{fig:FP-exponent}
	\end{figure}
	Thus, we can again identify a critical value $\lambda_c=1/2$ where $F(t)\sim t^{-1} (\log{t})^{-3/2}$ exhibits logarithmic corrections. Obviously, the first-passage exponent changes monotonically with the expansion parameter $\lambda$ (see Fig.~\ref{fig:FP-exponent}), whereas the dynamical exponent $z$ has been shown to stay constant for $\lambda\leq1/2$ (see Sec.~\ref{sec:CharacteristicsAlgebraical}). For the static case ($\lambda=0$) we recover the well-known exponent $\kappa=-3/2$ \cite{redner}. In the deflation case ($\lambda<0$), the first-passage rate declines faster when the deflation rate is increased, which is to be expected.
	
	\begin{figure}[t]
		\centering\includegraphics[width=0.47\textwidth]{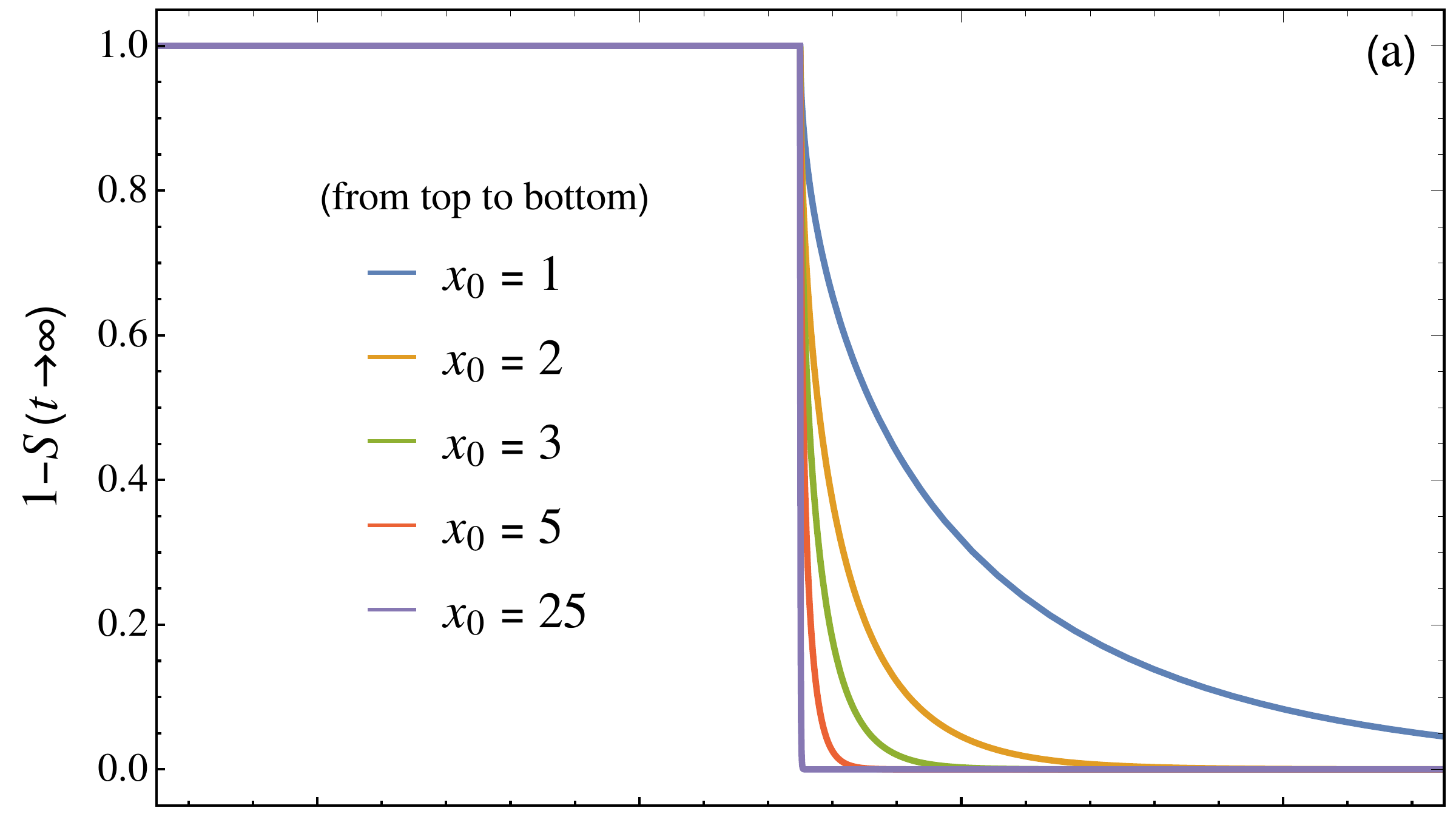}
		\hspace{0.6cm}
		\centering\includegraphics[width=0.47\textwidth]{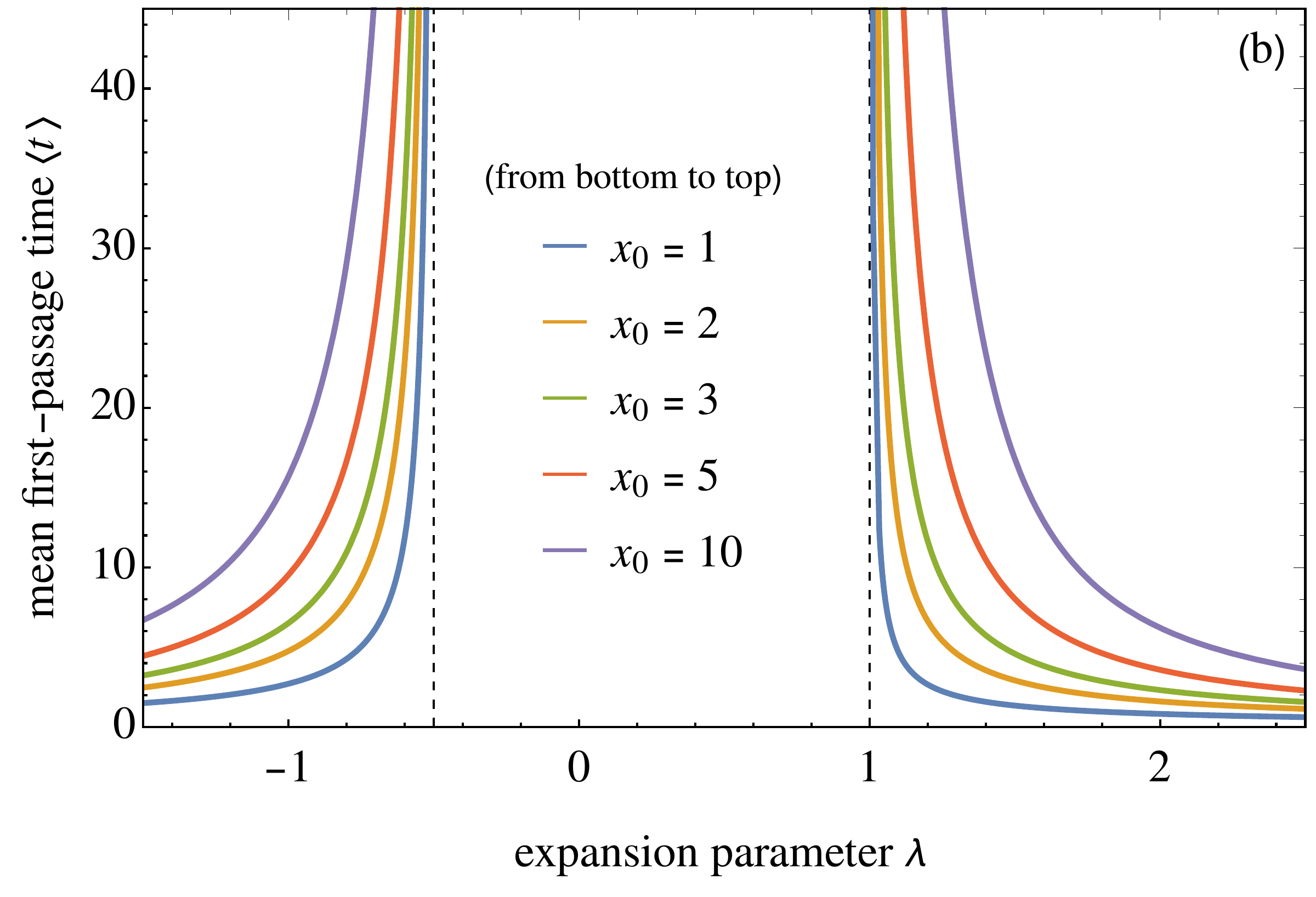}
		\caption{(Color online) Characteristics for algebraic dynamics. (a) Integrated first-passage probability ${1-S(t\rightarrow \infty)}$ for a diffusing particle with $D=1/2$ in an algebraically driven system as a function of the expansion parameter $\lambda$. (b) Corresponding mean first-passage time $\langle t \rangle$.}
		\label{fig:FPpow}
	\end{figure}	
	
	For positive values of $\lambda$ we have two regimes. The first-passage exponent rises for $\lambda<1/2$, exhibits a maximal value of $\kappa=-1$ for $\lambda_c=1/2$ and declines again for supercritical expansion ($\lambda>\lambda_c$). This behavior becomes clear if we consider the long-time limit of the survival probability $S(t)$, which accounts for all particles that have not yet passaged the origin by time $t$ and are therefore still part of the system. It is defined as $S(t)=\int_0^\infty \d x\,  c(x,t)$. Accordingly, $\int_{t_0}^t \d t' F(t')=1-S(t)$ specifies the amount of particles that already have passed the origin by time $t$. In our case, we get
	\begin{equation}
		S(t)=\mathrm{erf}\left[\frac{\langle x \rangle_{\lambda}}{\sqrt{2}\,\xi_\lambda} \right]=\mathrm{erf}\left[\frac{x_0 t^\lambda}{\sqrt{2}} \sqrt{\frac{1-2\lambda}{2D(t-t^{2\lambda})}}\right].
	\end{equation}
	The asymptotic limit
	\begin{multline}
		\lim_{t\rightarrow \infty} 1-S(t)\\=\left\{\begin{array}{cl}
			1 & \quad\mbox{ for}\quad \lambda\leq1/2, \\
			1-\mathrm{erf}\left[\frac{x_0}{\sqrt{2D}}\sqrt{\lambda-\frac{1}{2}}\right] & \quad\mbox{ for}\quad \lambda>1/2
		\end{array}\right.
		\label{eq:survivallimitPOW}
	\end{multline}
	describes the probability of a particle starting from a position $x_0>0$ to eventually reach the origin, see Fig.~\ref{fig:FPpow}(a). Note, that this result contains the recurrence property for a one-dimensional static space ($\lambda=0$), i.e.~the particle is certain to eventually passage the origin independent of its initial starting position. Of course, recurrence is still valid if the space is contracting since the particle is driven to the origin even faster. However, even if the underlying space is inflating, the particle will passage the origin with certainty as long as the expansion parameter $\lambda\leq1/2$. This feature is consistent with the previous observation that the particle exhibits ordinary diffusive behavior (dynamical exponent $z=2$) as inflation is too weak to have noticeable effect on the dynamics. However, as soon as $\lambda>1/2$, inflation dominates the overall process, recurrence does no longer apply and the system becomes \textit{transient}. In this case there exists a nonzero probability that a particle will eventually \emph{not} passage the origin. In the limit of an infinitely strong expansion $\lambda\rightarrow \infty$ the survival probability approaches unity since the particle has no chance to ever hit the origin. $S(t)$ becomes a step function in this limit (see Fig.~\ref{fig:FPpow}).
	
	The asymptotic behavior of the first-passage rate as a function of time, expressed in terms of the first-passage exponent $\kappa$, Eq.~\eqref{eq:kappa}, can be explained using the aforementioned properties. For $\lambda\leq 1/2$ the integral of $F(t)$ over time ${t\in(t_0,\infty)}$, i.e., the area under the curve equals one (due to the recurrence property). For small times, $F(t)$ is obviously larger for smaller values of $\lambda$, therefore, $F(t)$ declines faster with time, preserving normalization. For $\lambda>1/2$, however, recurrence does not apply and the area under the curve becomes smaller with larger $\lambda$.  Accordingly, the first-passage rate falls off faster because of the strong inflation of space. Taking both cases together, the first-passage exponent $\kappa$ exhibits a maximum at the critical value $\lambda_c=1/2$, i.e., the decline of $F(t)$ is slowest in this case.

	\begin{figure}[t]
		\centering\includegraphics[width=0.47\textwidth]{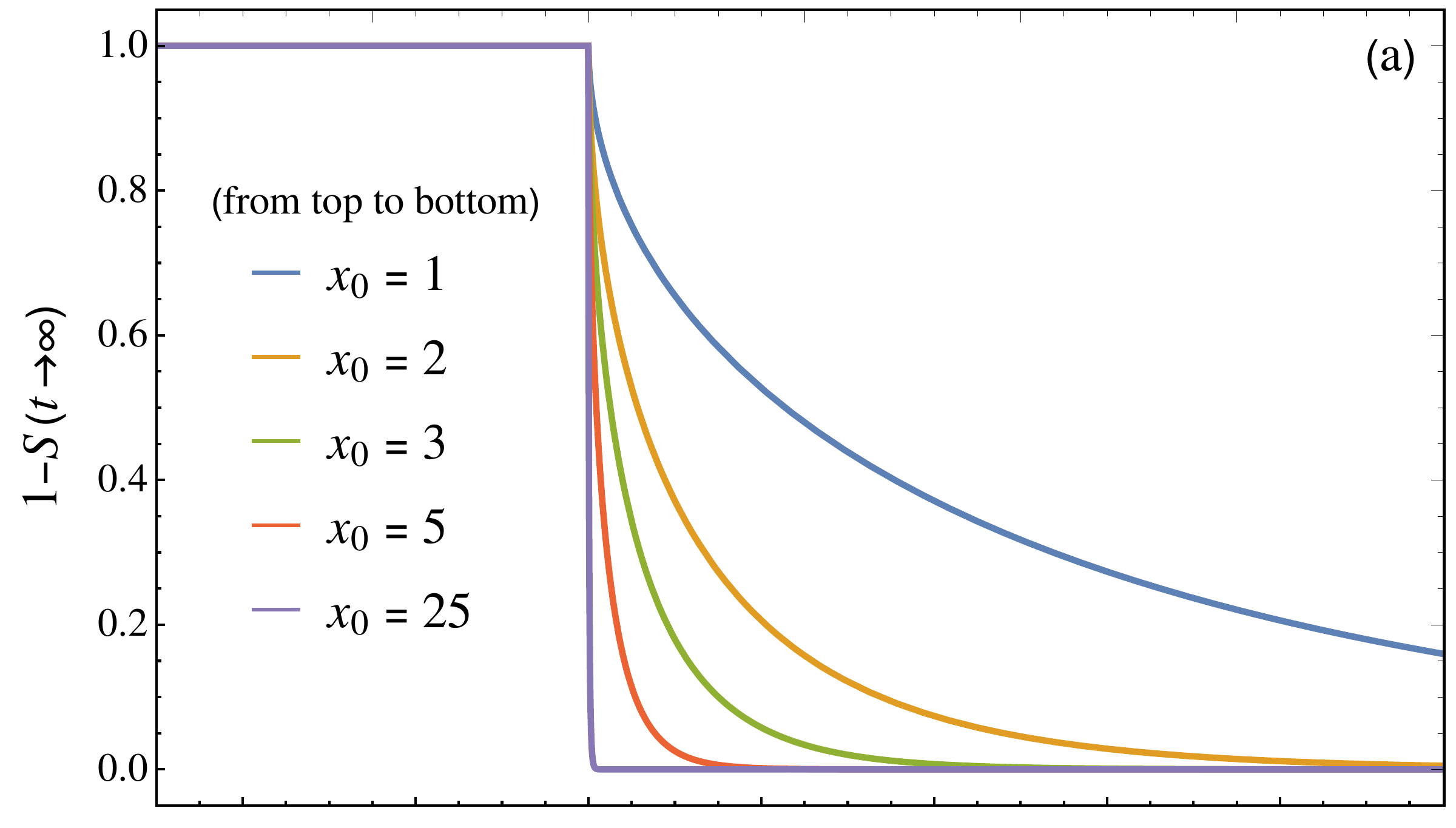}
		\hspace{0.6cm}
		\centering\includegraphics[width=0.47\textwidth]{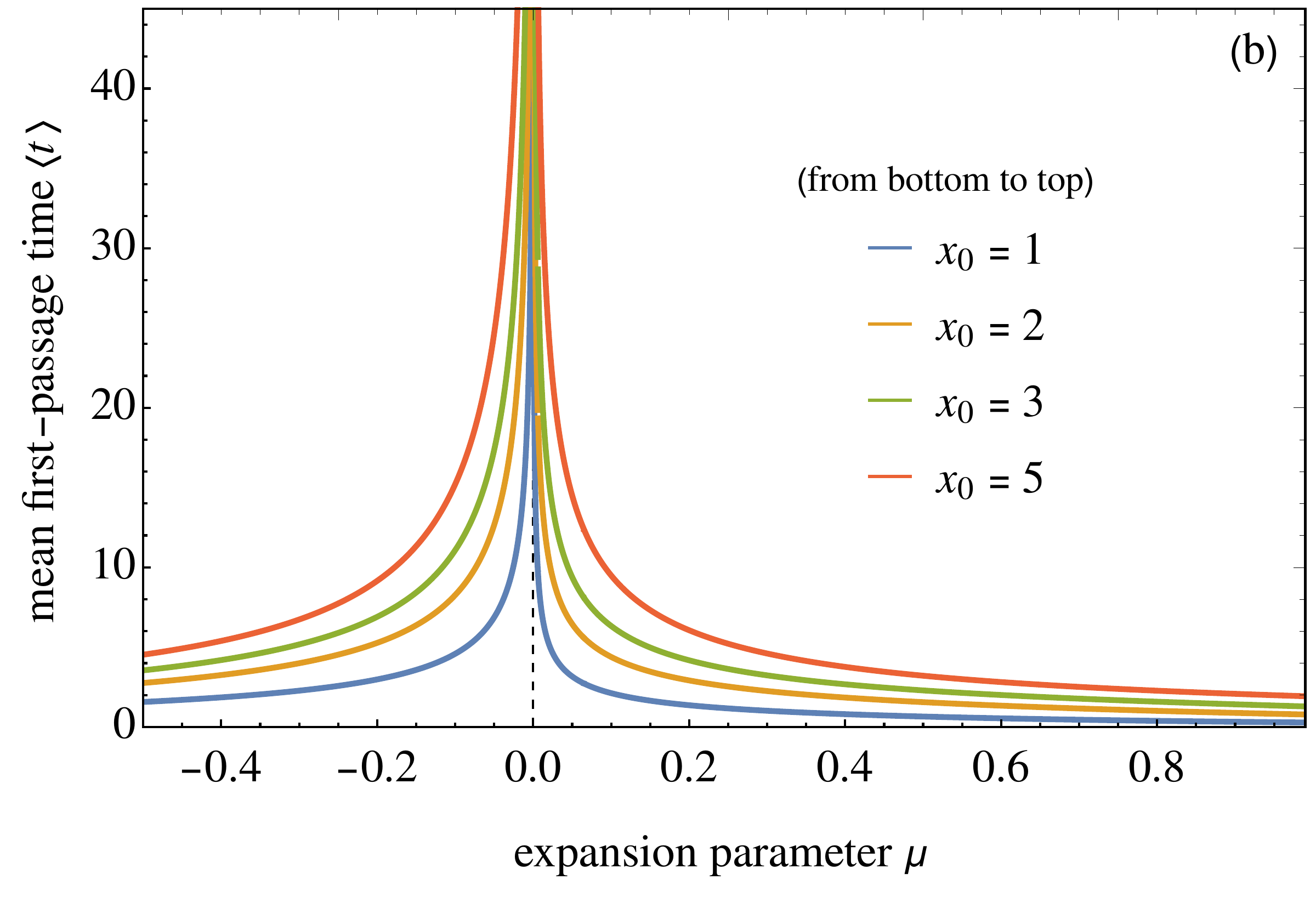}
		\caption{(Color online) Characteristics for exponential dynamics. (a) Integrated first-passage probability ${1-S(t\rightarrow \infty)}$ for a diffusing particle with $D=1/2$ in an exponentially driven system as a function of the expansion parameter $\mu$. (b) Corresponding mean first-passage time $\langle t \rangle$. }
		\label{fig:FPexp}
	\end{figure}
	
	Another interesting quantity is the first-passage time, defined as
	\begin{equation}
		\left<t\right>\equiv\frac{1}{1-S(t\rightarrow\infty)}\int_{t_0}^\infty\d t\,F(t)\,(t-t_0),
		\label{eq:MeanFirstPassageTime}
	\end{equation}
	which denotes the mean time that a particle takes to passage the origin for the first time. Numerical integrations of Eq.~\eqref{eq:MeanFirstPassageTime} are displayed in Fig.~\ref{fig:FPpow}(b). For the static case $(\lambda=0)$ we recover the known result that the mean time to pass the origin is infinitely large even though all particles eventually reach it. For infinitely strong contraction of space, i.e. $\lambda\rightarrow-\infty$, the mean time approaches zero since all particles are immediately trapped by the absorbing sink. Therefore, the first-passage time exhibits a finite value for strong contraction rates but approaches infinity for $\lambda\rightarrow -1/2$ where the diffusive spreading outweighs the contraction process. From the asymptotic behavior of the first-passage rate $F(t)\sim t^\kappa$ with exponent $\kappa$ given by Eq.~\eqref{eq:kappa} it follows that the first-passage time diverges as
	\begin{align*}
		\lim\limits_{\lambda\rightarrow -1/2^-}\langle t\rangle \sim \dfrac{1}{|\lambda|-\frac{1}{2}}
	\end{align*}
	In the interval $-1/2\leq \lambda \leq 1$, the particles on average travel an infinitely long time before arriving at the origin. As soon as $\lambda>1$, the first-passage rate declines faster than $t^{-2}$ which is why the mean time is again finite. This feature is called \textit{strong transience} \cite{jain1968}, meaning that $F(t)$ decreases so fast with the advance of time that a particle is able to passage the origin only within a finite amount of time. When approaching $\lambda\rightarrow 1$ from the right, the mean time diverges like $\frac{1}{2}(\lambda-1)^{-1}$. For a very large expansion of space the mean time $\langle t \rangle \rightarrow 0$. In this case, the survival probability is almost one and a particle can only be absorbed immediately after the start.
	
	Following the established term \emph{strong transience} for the region where the system is transient but $\langle t\rangle$ is finite, we term the effect of recurrence combined with finite mean return time as \emph{strong recurrence}. Fig.~\ref{fig:FP-exponent} shows a summary of the four qualitatively different regions.

	\subsection{Characteristics for exponential dynamics}
	\label{sec:FP_Exponential}
	
	For diffusing particles on an exponentially inflating or deflating support the first-passage rate scales as
	\begin{equation}
		\begin{aligned}
			F(t)&\sim \frac{e^{\mu t}}{\xi_{\mu}^3}\sim\left[\frac{\mu}{e^{\frac{4}{3}\mu t}-e^{-\frac{2}{3}\mu t}}\right]^{3/2}\\[1pt]
			\\&\sim\left\{\begin{array}{cl}
				|\mu|^{3/2} \,e^{-|\mu| t}& \quad\mbox{ for}\quad \mu<0, \\
				\mu^{3/2}\,e^{-2\mu t} & \quad\mbox{ for}\quad \mu>0,
			\end{array}\right.
			\label{eq:firstpassageEXP}
		\end{aligned}
	\end{equation}
	in the asymptotic limit. As a result of the non-algebraic behavior, $\kappa$ is formally infinite.
	
	Similar to the previous section, the integrated first-passage probability $1-S(t)$ can be calculated explicitly and yields in the long-time limit 
	\begin{equation}
		\lim_{t\rightarrow \infty} 1-S(t)=\left\{\begin{array}{cl}
			1 & \,\mbox{if}\quad \mu\leq0, \\
			1-\mathrm{erf}\left[\frac{x_0}{\sqrt{2D}}\sqrt{\mu}\right] & \,\mbox{if}\quad \mu>0.
		\end{array}\right.
	\end{equation}

	Accordingly, the particle is only recurrent if the space is static or contracting, which is consistent with the observation that the behavior of the system is still diffusive in these cases. However, as soon as we switch on expansion with any (arbitrarily small) value $\mu>0$ the behavior of the system becomes transient. Following the same line of reasoning as in the algebraic case, it is plausible that the first-passage rate \eqref{eq:firstpassageEXP} declines more strongly for expansion than in the case of contraction of space.
	
	The mean first-passage time $\langle t \rangle$ is again integrated numerically and shown in Fig.~\ref{fig:FPexp}(b). It stays finite for $\mu\neq 0$ and diverges $\sim|\mu|^{-1/2}$, but with different amplitudes on either side, in the limit $\mu\rightarrow 0$ from the left and from the right, respectively. As a consequence, we have only three regions here, namely strong transient behavior for $\mu>0$, recurrent behavior for $\mu=0$ and strong recurrent behavior for $\mu<0$.

	\subsection{Characteristics for exponential dynamics in multidimensional spaces}
	
	We are also interested in the first-passage properties of diffusion in higher-dimensional expanding or contracting systems. The absorbing sink is in this case given by a spherical shell with radius ${r=R}$ centered at the origin and the diffusing particle is initially starting from a position outside the shell with a given radius ${r=r_0>R}$. Formally, the concentration $c(r,t)$ of the particle in this spherically symmetric setting is directed by the radial inflation-diffusion equation \eqref{eq:sphericalFokker} with boundary condition $c(r=R,t)=0$ and with initial condition given by Eq.~\eqref{eq:RadialInitialCondition}.
	
	Unfortunately, deriving an exact solution of $c(r,t)$ is not as simple as in the one-dimensional problem. In particular, the image method is not applicable since the probability density \eqref{eq:pdf_HD} is invariant under a parity transformation ${r\rightarrow -r}$. Using instead an ansatz with inverse radius transformation fulfills the boundary conditions, however, it does not solve Eq.~\eqref{eq:pdf_HD}. Nonetheless, we can at least derive first-passage properties for a time-independent Hubble parameter $\mu$, i.e., the exponential case by solving the Laplace transformed inflation-diffusion equation
	\begin{multline*}
		\frac{\delta(r-r_0)}{\Omega_d\,r_0^{d-1}}-s\,\tilde{c}(r,s)=\\
		\frac{\mu}{r^{d-1}}\frac{\partial}{\partial r}\left[r^d \tilde{c}(r,s)\right] -D\,\frac{1}{r^{d-1}}\frac{\partial}{\partial r}\left[r^{d-1}\frac{\partial}{\partial r}\tilde{c}(r,s)\right]
		\label{eq:FokkerLaplace}
	\end{multline*}
	with concentration ${\tilde{c}(r,s)=\int_0^\infty \d t\, c(r,t)\,e^{-s\,t}}$. From the transformed first-passage rate
	\begin{equation}
		\tilde{F}(s)=D\,\Omega_d\,R^{d-1}\,\frac{\partial}{\partial r}\tilde{c}(r,s)\Big|_{r=R}.
	\end{equation}
	the long-time limit of the survival probability
	\begin{equation}
		S(t\rightarrow\infty)=1-\tilde{F}(s\rightarrow 0),
	\end{equation}
	as well as the mean first-passage time
	\begin{equation}
		\left<t\right>=-\frac{\partial}{\partial s}\log \tilde{F}(s)\Bigg|_{s\rightarrow 0},
		\label{eq:meantimeHD}
	\end{equation}
	can be calculated.
	
\begin{figure}[t]
		\centering\includegraphics[width=0.47\textwidth]{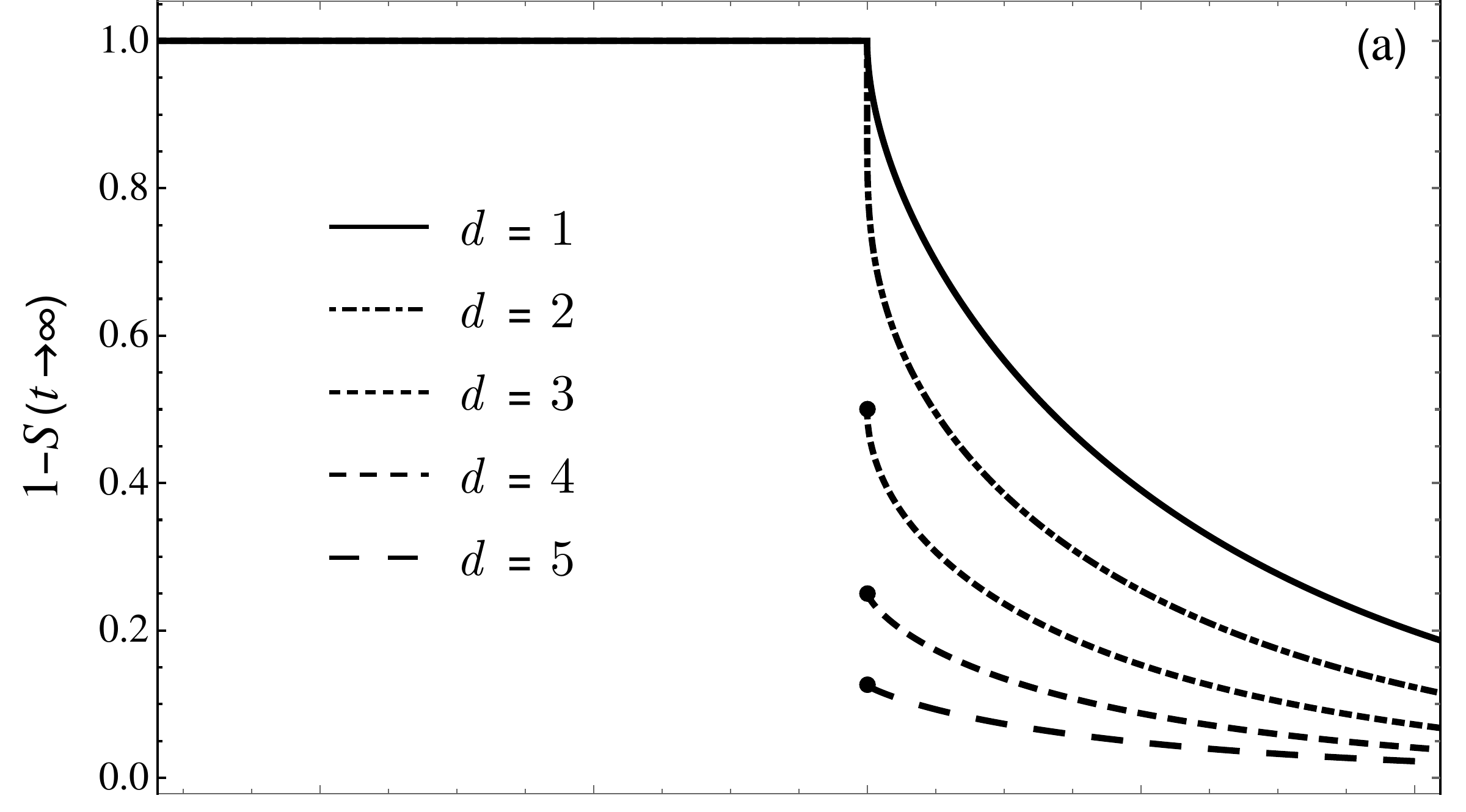}
		\hspace{0.6cm}
		\centering\includegraphics[width=0.47\textwidth]{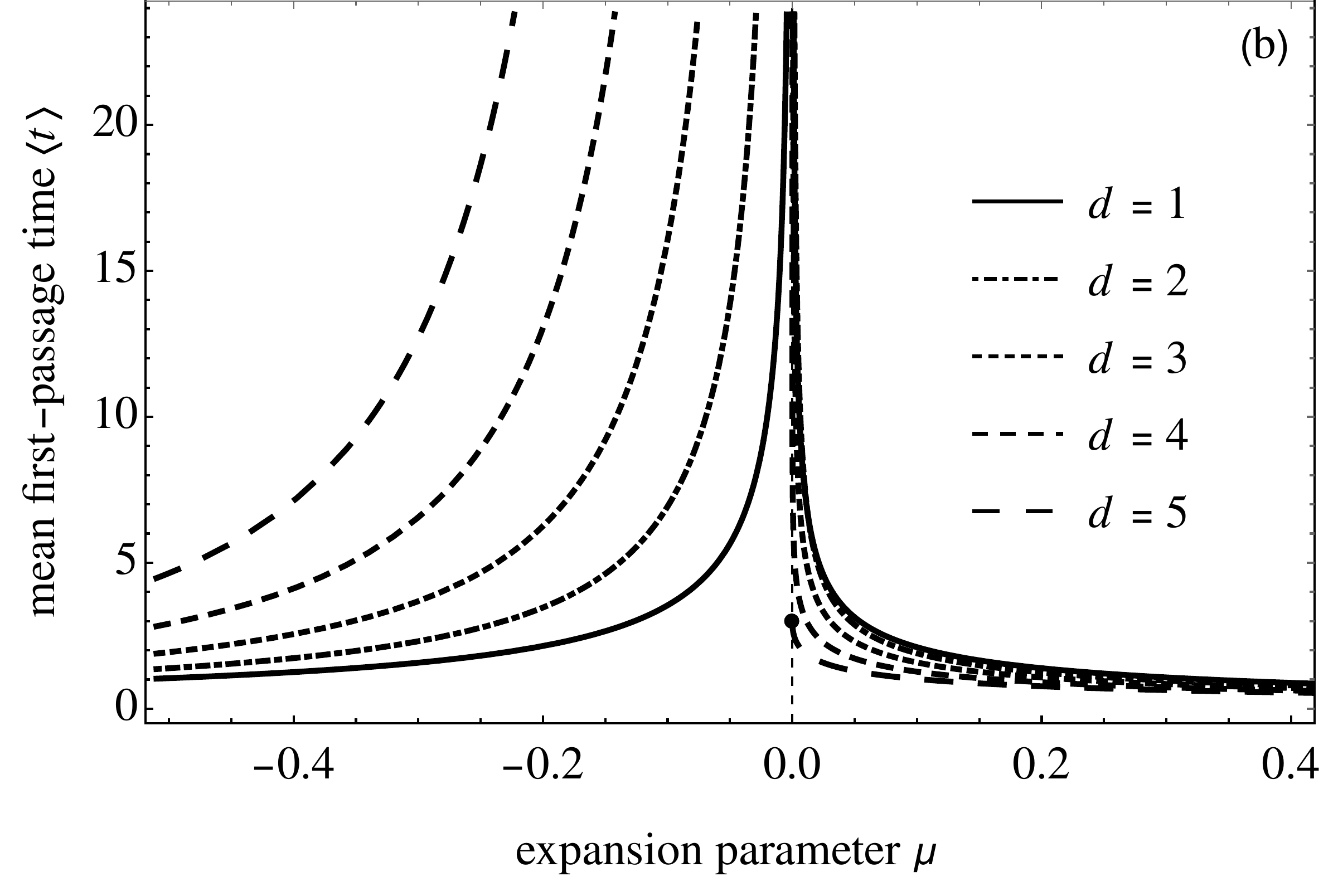}
		\caption{(a) Integrated first-passage probability according to \eqref{eq:survivalHD} for a particle in $d$-dimensional space with parameters $R=1$, $r_0=2$, and $D=1/2$. (b) Corresponding mean first-passage time $\left \langle t\right \rangle$. Note that $\left\langle t \right \rangle$ is infinitely large for an expansion parameter $\mu=0$ if $d\leq4$, whereas it is finite otherwise.}
		\label{fig:FPdim}
	\end{figure}

	The explicit calculation of $\tilde{c}(s,t)$ and $\tilde{F}(s)$ is omitted as the expressions turn out to be quite cumbersome and are not instructive for further discussions. Eventually, the integrated first-passage probability reads
	\begin{equation}
		1-S(t\rightarrow \infty)=\left\{\begin{array}{cl}
			1 & \quad\mbox{ for}\quad \mu<0, \\
			\frac{\Gamma \left(1-\frac{d}{2},\frac{r_0^2 \mu }{2 D}\right)}{\Gamma \left(1-\frac{d}{2},\frac{R^2 \mu }{2 D}\right)} & \quad\mbox{ for}\quad \mu\geq0,
		\end{array}\right.
		\label{eq:survivalHD}
	\end{equation}
	where $\Gamma(a,b)$ denotes the \textit{incomplete gamma function}. It is plotted in Fig.~\ref{fig:FPdim}(a) for various dimensions $d$. In Fig.~\ref{fig:FPdim}(b), the numerical evaluation of the mean first-passage time is shown.
	In the deflation case, the particle shows strong recurrent behavior in all dimensions since $ 1-S(t\rightarrow \infty) = 1 $ and $ \langle t\rangle < \infty $  for all $\mu<0$. For increasing dimension $d$ the mean time increases since the particle has more chances to skirt the absorbing sink before trapping takes place. By taking the limit of \eqref{eq:survivalHD} for $\mu\rightarrow 0$ we obtain
	\begin{equation}
		1-S(t\rightarrow \infty)=\left\{\begin{array}{cl}
			1 & \quad\mbox{ for}\quad d\leq2, \\
			\left(\frac{R}{r_0}\right)^{d-2} & \quad\mbox{ for}\quad d>2
		\end{array}\right.
		\label{eq:survivalHDstatic}
	\end{equation}
	in accordance with the findings presented in \cite{redner} for the static case. Evidently, ${d=2}$ denotes a critical dimension where the recurrent behavior for ${d\leq2}$ changes into transient behavior for $d>2$. Therefore, the curves of $1-S(t\rightarrow \infty)$ are continuous for $d\leq2$ but only right-continuous at $\mu=0$ for $d>2$. For dimensions $d>4$ the mean first-passage time is finite, i.e., the particle is strongly transient even in the static case. For exponentially driven expansion however, we find strong transience for any dimension, as was to be expected from the one-dimensional results in Sec.~\ref{sec:FP_Exponential}. For increasing dimension the integrated first-passage probability decreases since the particle is less likely to eventually passage the absorbing shell. Finally, for very strong expansion, both $ 1-S(t\rightarrow \infty) $ and $ \langle t\rangle$ rapidly decay to zero for any dimension which is also consistent with the one-dimensional results.

\section{Summary}
\label{sec:Summary}
	We derive a generalized diffusion equation for an underlying support that is evolving in time. For the two particular cases of an exponential or algebraic expansion and contraction we are able to solve this equation in a closed form. Given the algebraic dynamics, diffusion and background evolution act as competing processes. This fact allows us to calculate the dynamical exponent as a function of the expansion parameter $ \lambda $. For exponential dynamics, governed by a different parameter $ \mu $, however, the overall process is always dominated by expansion, but as soon as the support is contracting, we get a stationary solution only depending on the two time-scales involved. 
	
	Moreover, we derive first-passage properties for our setting analytically. Most interesting is the first-passage exponent, which in the algebraic case, turns out to vary with~$\lambda$. Considering also the mean first-passage time, we find four extended qualitatively different dynamical regimes, where the behavior is either strong recurrent, recurrent, transient or strong transient. In the case of exponentially driven expansion/contraction (tuned by an inverse time-scale $ \mu $), no exponents can be defined and only three regimes are found. Finally, we also derive analytic expressions for the exponential case in more than one dimension and show that our results are compatible with the known behavior in the static limit ($ \mu = 0 $).

\begin{acknowledgments}
	We thank H.~Hinrichsen, P.~Fries and J.~S.~E.~Portela for useful discussions. M.~Schrauth acknowledges the financial support from the Studienstiftung des deutschen Volkes. This work is part of the DFG research project Hi~744/9-1. 
\end{acknowledgments}

\end{document}